\documentclass{eptcs}  
\providecommand{\event}{QAPL 2015} 
\usepackage[utf8]{inputenc}
\usepackage{graphicx} 
\usepackage{amsmath}
\usepackage[table]{xcolor}
\usepackage{textcomp}
\usepackage{hyperref} 

\newcommand{\entra}{\thanks{%
 The research leading to these results has received funding from
 the European Union Seventh Framework Programme (FP7/2007-2013)
 under grant agreement no 318337,
 ENTRA - Whole-Systems Energy Transparency.}}

\let\oldvec=\vec
\def\vec#1{\oldvec{\,#1}}

\def\paragraph#1{\par\medskip\noindent{\textbf{#1.}} }
\newtheorem{definition}{Definition}
\newtheorem{lemma}{Lemma}
\def\bbbr{I\!\!R}
\def\enddot{\hspace*{1mm}\mbox{.}}

\def\kw#1{\;\mbox{\textbf{#1}}\;}
\def\bc#1{\langle#1\rangle}
\def\Ppup{\overline{P}_{\mathtt{p}}}
\def\Ppdwn{\underline{P}_{\mathtt{p}}}
\def\Fup{\overline{F}} 
\def\Fdwn{\underline{F}}
\newcommand{\etal}{{et al.}}

\def\mspaceend{}
\def\mspacebegin{}

\begin{document}\sloppy%
\title{Probabilistic Output Analysis by Program Manipulation} 
\author{Mads Rosendahl\quad\quad Maja H. Kirkeby\entra\institute{Computer Science, Roskilde University, Denmark}\email{ $\{$madsr,majaht$\}$@ruc.dk}}
\def\titlerunning{Probabilistic Output Analysis by Program Manipulation}
\def\authorrunning{Mads Rosendahl {\&} Maja H. Kirkeby}
\maketitle


\begin{abstract}

The aim of a probabilistic output analysis is to derive a probability distribution of possible output values for a program from a probability distribution of its input. 
We present a method for performing static output analysis, based on program transformation techniques. 
It generates a probability function as a possibly uncomputable expression in an intermediate language. This program is then analyzed, transformed, and approximated. 
The result is a closed form expression that computes an over approximation of the output probability distribution for the program. 
We focus on programs where the possible input follows a known probability distribution. 
Tests in programs are not assumed to satisfy the Markov property of having fixed branching probabilities independently of previous history.
\end{abstract}


\section{Introduction}
\label{sec:Introduction}

The aim of a probabilistic output analysis (POA) is to derive a probability distribution for output values from a probability distribution for input to a program. 
Internal properties of a program can also be analyzed in this way by instrumenting programs with step-counters for complexity analysis
\cite{conf/fpca/Rosendahl89} or energy consumption measures 
\cite{conf:lopstr:Liqat2013}. 

When analyzing energy consumption, probability distributions may provide more useful information than boundaries. 
Wierman et al. states that \textit{``global energy consumption is affected by the average case, rather than the worst case``} \cite{conf:allerton:Wierman2008}. 
Also in scheduling \textit{``an accurate measurement of a tasks average-case execution time (ACET) can assist in the calculation of more appropriate deadlines''} \cite{jour:cc:Guo2007}. 
For a subset of programs a precise average case execution time can be found using static analysis \cite{journals/tcs/FlajoletSZ91,thesis:cork:gao2013,books/daglib/0020847}. 
In some cases the POA delivers not only an accurate output average but the more descriptive accurate output distribution. 
In other cases the POA must over approximate the probability distribution and the expected value (average case result) will be approximated safely as a range. 
Another application area for POA is in temperature management, where worst-case bounds are important \cite{conf/rtas/SchorBYT12}. 
Because POA return distributions it can be used to calculate the probability of energy consumptions above a certain limit, and thereby indicating the risk of over-heating. 

The main contribution in this paper is to present a technique for probabilistic analysis where the analysis is seen as a program-to-program translation. 
This means that the transformation to closed form is a source code program transformation problem and not specific to the analysis. Any necessary approximation in the analysis is also performed at the source code level.
The technique also makes it possible to 
balance the precision of the analysis against the brevity of the result.

The method in this paper is inspired by the techniques used in automatic complexity analysis. 
Wegbreit's Metric system \cite{journals/cacm/Wegbreit75} laid the ground work for many later systems with an aim of deriving least, worst and average case complexity measures. 
Later works in this area have focused on worst case complexity \cite{journals/entcs/AlbertAGP09,conf/iclp/Lopez-GarciaDB10,conf/fpca/Rosendahl89} with advanced systems that can analyze realistic programs. 
The approach in this paper uses an approach similar to automatic complexity analysis \cite{conf/fpca/Rosendahl89} in that we derive the probability distribution without approximations but only in the last phase introduce approximations. 
We use a simple first-order functional language with restricted recursion for the analysis, but it can be seen as an intermediate language for analysis of programs in other languages.
We transform the original program into a program that computes the probability distribution and this program can then be analyzed, transformed, and approximated. 
It is thus an alternative to deriving cost relations directly from the program
\cite{journals/entcs/AlbertAGP09,conf/iclp/Lopez-GarciaDB10} or expressing costs as abstract values in a semantics for the language.

As with automatic complexity analysis the aim of probabilistic output analysis is to extract the result as a parameterized expression. 
The time complexity of a program should be stated as a closed form expression in the input size and for probabilistic output analysis the aim is to find the probability of output values of the program as a function in output values and input size or range.
As a small example let us consider the addition 
\texttt{add}
of two independent integer values  \texttt{x} and \texttt{y} evenly distributed from 1 to $n$. 
It is a tail-recursive program where the output distribution is well-known to be a triangular shaped distribution. The program and the input probability distributions should both be expressed in the same language as they are part of the transformational approach to obtain the output probability distribution.
\begin{quote}\vskip 2pt
\begin{verbatim}
add(x,y) = if (x=0) then y else add(x-1,y+1)
px(x,n) = if (x >= 1 and x <= n) then 1/n else 0
py(y,n) = if (y >= 1 and y <= n) then 1/n else 0
\end{verbatim}\vskip 2pt
\end{quote}
Our probabilistic output analysis returns a function describing the probability distribution of the output:
\begin{quote}\vskip 2pt
\begin{verbatim}
padd(z,n) = 
  1/(n*n)*max(min(n,z-1) - max(1,z-n) + 1,0)
\end{verbatim}\vskip 2pt
\end{quote}
The probability distribution is here a closed form expression parameterized in the output value ({\tt z}) and range of input values ({\tt n}).
The analysis can also be used for more complex input distributions and programs but it will not always be able to reduce it to a precise result in closed form. 
If this is not possible we will approximate the distribution and thus get an over approximation of the extreme cases and a range for the expected value.
If input values are not independent we can specify a joint distribution for the values.


\section{Probability distributions}

The analysis presented here assumes a discrete set of values for input and output. 
The set will be finite or countable and we will use discrete probability distributions.
We consider the input to a program as a discrete random variable and the input probability distribution is then a probability measure that to an event of input having a given value assigns a value between 0 and 1. 
This is also often referred to as the \emph{probability mass function} in the discrete case, and in the continuous case the equivalent is the \emph{probability density functions}. 
We will use the phrase \emph{probability distribution} to denote mappings from single values (input or output) to a probability or number between 0 and 1. Distribution will be denoted with an upper case $P$ letter.

\begin{definition}[input probability]\label{def:inputDist}
For a countable set $X$ an input probability distribution is a mapping $P_x:X\rightarrow \{r\in\bbbr\mid 0\leq r\leq 1\}$, where
\begin{align*}
 \sum_{x\in X} P_x(x) = 1  
\enddot
\end{align*}
\end{definition}
We define the output probability distribution for a program {\tt p} in a forward manner. 
It is the {\em weight} or sum of all probabilities of input values where the program returns the desired value $z$ as output.

\begin{definition}[output probability]\label{def:outputDist}
 Given a program, $\mathtt{p} : X \rightarrow Z$ and a probability distribution for the input,  $P_{X}$, the output probability distribution, $P_{\mathtt{p}}(z)$, is defined as:
 \begin{align*}\mspacebegin
  P_{\mathtt{p}}(z) & = \sum_{x \in X \land \mathtt{p}(x) = z} P_{X}(x) 
\enddot
 \end{align*}
\end{definition}
Note that Kozen also uses a similar forward definition 
\cite{journals/jcss/Kozen81}, 
whereas Monniaux constructs the inverse and expresses the relationship in a backwards style  \cite{conf/sas/Monniaux00}. 
\begin{lemma}\label{lem:sumOutBound}
The output probability distribution, $P_{\mathtt{p}}(z)$, 
satisfies
\begin{align*}\mspacebegin
 0 \leq \sum_{z} P_{\mathtt{p}}(z) \leq 1
\enddot
\mspaceend
 \end{align*}
\end{lemma}
The program may not terminate for all input and therefore the sum may be less than one. 
If we expand the domain $Z$ with an element to denote non-termination, $Z_\bot$, the total sum of the output distribution $P_{\mathtt{p}}(z)$ would be 1. 

\paragraph{Approximations of probability distributions}
The output analysis cannot necessarily derive the precise probability distribution. 
Various approaches to approximations of probability distributions have been proposed and can be interpreted as {\em imprecise probabilities}
\cite{conf:vstte:AdjeB2014,conf:isip:Destercke2009,tech:compegne:Ferson2014}. 
Dempster-Shafer structures \cite{conf:rbes:Gordon1984,conf/uai/Bauer96} 
and P-boxes \cite{conf:isip:Destercke2009,tech:sand:Ferson2002} can be used to capture and propagate uncertainties of probability distributions. 
There are several results on extending arithmetic operations to probability distributions for both known dependencies between random variables and when the dependency is unknown or only partially known
\cite{journals/rc/BerleantC98,journals/computing/BouissouGGP12,conf:rta:Kay2007,thesis:Stellenbosch:Uwimbabazi2013,conf:hdrums:Wilson2000}. 
Algorithms for lifting basic operations on numbers to basic operations on probability distributions can be used as abstractions in static analysis based on abstract interpretation \cite{conf/sas/Monniaux00}.
Our approach uses the P-boxes as bounds of probability distributions.
P-boxes are normally expressed in terms of the cumulative probability function but we will here use the probability mass function. We do not, however, use the various basic operations on P-boxes but apply approximations to a probability program such that it forms a P-box.

\begin{definition}[over and under approximation]\label{def:approximationP}
For a distribution $P_{\mathtt{p}}$ an over and under approximation 
 ($ \Ppup{}$ and $ \Ppdwn{}$)
of the distribution satisfies the conditions:
 \begin{align*}\mspacebegin
  & \Ppup{} :  \forall z . 
    P_{\mathtt{p}}(z) \leq  \Ppup{}(z) \leq 1
\\[\smallskipamount]
  &
   \Ppdwn{} :  \forall z . 
      0 \leq  \Ppdwn{}(z) \leq P_{\mathtt{p}}(z) 
\enddot
\mspaceend      
 \end{align*}
\end{definition}
The aim of the output analysis is to derive as tight approximations  $ \Ppdwn{}$ and $ \Ppup{}$ as possible. 

\begin{lemma}\label{lem:sumPbounds}
Given the definition for over and under approximation 
they will have boundaries for their total weights as 
\begin{align*}\mspacebegin
& 0 \leq \sum_{z}  \Ppup{}(z) \leq \infty 
\quad\quad
0 \leq \sum_{z}  \Ppdwn{}(z) \leq 1
\enddot
 \end{align*}
\end{lemma}
When $ \Ppdwn{} =  \Ppup{}$ the total weight for each function will be equal to the total weight of $P_{\mathtt{p}}$, according to definition \ref{def:approximationP}. 
For terminating programs the total weight is 1.

\paragraph{Expected value}\label{sec:ExpectedValue}
Provided that the output from the program is numerical, one may be interested in the average output value of the program or \emph{the expected value} of the output distribution. 
If the program does not terminate for all input it is not clear how to define the expected value so as part of the further analysis we need a guarantee that the program terminates. 
If the sum of the $ \Ppdwn{}$ is $1$ then we know that the program terminates for all possible input (\emph{i.e.} input with probability greater than zero). 

\begin{lemma}\label{lem:flatterm}
The under approximation of a probability distribution satisfies
\begin{align*}\mspacebegin
 \sum_{z}  \Ppdwn{}(z) = 1
 \Rightarrow
 \sum_{z} P_{\mathtt{p}}(z) = 1
\enddot
\mspaceend
\end{align*}
\end{lemma}
A similar implication does not apply to the over approximation.
The expected value of the output distribution is defined as the weighted average of the distribution.
\begin{definition}[expected value]
The expected value of the output distribution is defined as
\begin{align*}\mspacebegin
E_{\mathtt{p}} = \sum_z z \cdot P_{\mathtt{p}}(z)
\enddot
\mspaceend
\end{align*}
\end{definition}
If we cannot analyze the program precisely, we can use the over approximation to compute  an interval for the expected value.
We cannot use the approximation $\Ppup{}$ directly as its weight is not necessarily 1. 
It can, however, be used to define over and under approximations to the cumulative probability distribution.
These two can then be used to calculate a lower and an upper bound for the expected values.
\begin{definition}[expected value interval]
For an over approximation of a probability distribution 
$ \Ppup{}$ we define an over and under accumulation 
($F^\uparrow$ and $F^\downarrow$)
and over and under expected value
($E^\uparrow$ and $E^\downarrow$).
\begin{align*}
&F^\uparrow(z) = \min(\sum_{v\leq z}  \Ppup{}(v),1)
&
\\
&F^\downarrow(z) = \max(1-\sum_{v\geq z}  \Ppup{}(v),0)
\\[\smallskipamount]
&E^{\downarrow} =
\sum_z z \cdot (F^\uparrow(z) - F^\uparrow(dec(z)))
&
\\
&E^{\uparrow} =
\sum_z z \cdot (F^\downarrow(z) - F^\downarrow(dec(z)))
\\[\smallskipamount]
&dec(z) = \max\{v \in Z \mid v < z\}
&&
\enddot
\mspaceend
\end{align*}
\end{definition}
Notice that an expected value based on an over approximation of the 
accumulated probability gives an under approximation of the expected value.
If the output space $Z$ is integers then the {\it dec} function will just subtract one from its argument.

\begin{lemma}[expected value interval]
For a terminating program the expected value can be approximated by an interval from the over approximation of the probability distribution.
\begin{align*}\mspacebegin
E^{\downarrow}\leq 
E_{\mathtt{p}}
\leq
E^{\uparrow}
\enddot
\end{align*}
\end{lemma}

\paragraph{Externalize resource usage}
The output analysis can be used to analyze internal properties of the program provided these properties are externalized. 
As in automatic worst case complexity analysis
\cite{conf/fpca/Rosendahl89}, this may be done by instrumenting the program with step counting information. 
Similarly we might instrument programs with energy consumption based on low level energy models for operations
\cite{conf:lopstr:Liqat2013} to be able to analyze programs 
for average energy consumption.

Usually an operational or denotational semantics of a simple first order functional programming language describes programs as mappings from input values to output values. 
The time, space or energy required to perform the computation would normally not be part of the semantics. 
A simple form of resource analysis is to count the number of basic operations that a computation would require. 
An automatic complexity analysis \cite{conf/fpca/Rosendahl89} is then based on a semantics that has been extended (or instrumented) with step-counting information so that the meaning of a program is a mapping from input values  to a tuple of the number of steps and the output value. 
If we write this semantics as an interpreter in the source language we can convert a program to a step-counting version of the program by partial evaluation. 
In this way the complexity analysis has been transformed into an output analysis of the program. The aim of the complexity analysis is then to generate an over approximation of the (first component of the) possible output as a function of the size of the possible input.
If the semantics is instrumented with other types of resource information we can analyze programs with respect to these properties.
Some automatic complexity analysis systems are based on translating programs into cost relations  \cite{journals/entcs/AlbertAGP09} or 
cost equations \cite{conf/iclp/Lopez-GarciaDB10}. These approaches are then used as approximations of an instrumented semantics that captures the cost of computations.

The functional language we use here may be seen as a meta language for the analysis since we should extend source programs with resource information before they are analyzed. One can also use it as a meta language for analyzing programs in other languages. This can be achieved by having a step-counting interpreter for the other language written in the first order functional language. When analyzing software for embedded systems the programs are often written in simple c-like languages which should then be translated into this meta language.

\paragraph{The challenge of approximation} 
Analysis of probabilistic behavior introduces some new challenges compared to worst case analysis. 
It is well known that a function of expected values is not necessarily the same as the expected value of the function.  
There are a number of other potential pitfalls when making approximations in a probabilistic setting.  One might assume that conditions in a program can be assigned a fixed probability of being true independently of previous execution paths in the program.
One might also assume that variables have independent probability distributions. 
An unfortunate effect of using independence as an approximation is that it tends to under approximate the extreme cases. 
In a throw of two dice the sum of $12$ has probability $1/36$ if we can assume independence. 
If (by some magic) they always showed the same the probability increases to $1/6$. 
The situation is well-known in the insurance industry 
and for financial risk management (valuation of derivatives)
where one may want to over approximate the risk of extreme event when events are not guaranteed to be independent. 
One approach to handle such situations is the use of copulas 
\cite{journals/jec/BernatBN05}
and  comonotonicity of probability  measures
\cite{journals:ima:DhaeneDGKV02}.


\section{Transformation Based Analysis}
\label{sec:OurTech}

Our analysis is based on a small first order functional language with primitive recursion.
The first step of the analysis is to translate programs into a new language of probability distribution programs.
We then use analysis and transformation techniques to transform the probability distributions into closed form. 
Failing that, we may over approximate the distribution  as discussed in section \ref{sec:ApproxTec}.

Programs ({\tt p}) are defined as a collection of functions
\begin{align*}
& f_1(x_1,\ldots,x_n) = e_1\\
&\vdots\\
&f_n(x_1,\ldots,x_n) = e_n  \quad\enddot
\end{align*}
The first function in the program is called externally and for 
that function we have an input probability distribution $P_x$ specified as a symbolic expression $e_x$.
The language uses a base set $D$ of values for simple expressions, and functions in a program denote mappings from tuples of values to values $D^* \rightarrow D$. 
The base set of values will not be further restricted here, nor do we specify the exact set of basic operations in the language.
Functions are either non-recursive or primitive recursive. The latter will have the form: 

\begin{align*}
&f(x_1,\ldots,x_n) = \kw{if}(b(x_1,\ldots,x_n)) \kw{then} g(x_1,\ldots,x_n) \kw{else} f(e_1\ldots,e_n)
\end{align*}
Non-recursive functions have right hand sides that are built from simple operations conditional expressions and function calls to non-recursive and primitive recursive functions.

\subsection{Probability distribution program} 
The output distribution program is expressed in a language similar to the original program but extended with an extra class of functions. It contains the original functions of type $D^* \rightarrow D$ and probability functions of type $D^* \rightarrow [0,1]$. One of these functions will be the output distribution function of type $D\rightarrow[0,1]$.
The language for probability distribution program uses two new language constructs: Sums  over the (possibly) infinite set of all input values in $D$ and a constraint function $C$. 
The constraint function eases the handling of boundaries and is defined as
\begin{align*}
C(condition) = \left\{ 
\begin{array}{ll} 1 & \textrm{ if } condition = true \\ 0 & 
\textrm{ otherwise } \\ 
\end{array} \right.
\end{align*}
This definition is related to the indicator function \cite{journals/toplas/MorganMS96} or characteristic function for membership of sets.
We also extend the language with a finite product construction which will be used for unfolding primitive recursion.

\smallskip

The first phase constructs the probability distribution program from the original program  and a joint input distribution function. The input distribution is defined as a function
\begin{align*}
&P_x(x_1,\ldots,x_n) = e_x 
\end{align*}
that computes the probability of each possible input value. If the arguments are independent the function can be written as a product of the probability distribution of each parameter
\begin{align*}
&P_x(x_1,\ldots,x_n) = P_{x1}(x_1)\cdots P_{xn}(x_n)
\end{align*}
The raw form of the probability distribution program is defined as follows.
Given an output value, the distribution program sums the probabilities for all input value tuples that the original program maps to the output value. 
The probability distribution program ($P_{\mathrm p}$) is defined as follows.
\begin{align*}
&P_{\mathrm p}(z) = \sum_{x_1} \cdots \sum_{x_n}P_x(x_1,\ldots,x_n)\cdot C(z=f_1(x_1,\ldots,x_n))
\\
&P_x(x_1,\ldots,x_n) = e_x 
\\[0.5\smallskipamount]
&f_1(x_1,\ldots,x_n)=e_1\\
&\vdots\\
&f_n(x_1,\ldots,x_n)=e_n
\end{align*}
We view a probability distribution program as a program that can be transformed and analyzed.
In the next phase function calls are unfolded 
and in the following phase the result is simplified using rewrite rules.

\subsection{Unfolding}
In this phase we unfold function calls in the program. We will introduce the central transformation rules for unfolding calls to functions in the original program based on the syntactical structure.

\paragraph{Function calls}
Simple calls to functions can be unfolded directly. Calls to primitive recursive function can be composed but each call can be analyzed separately by constructing a joint input distribution function to the call.
For such function calls we rewrite the program as follows
\begin{align*}
\mspacebegin
&\sum_{x_1}\cdots\sum_{x_n} P_x(x_1,\ldots,x_n) \cdot C(z= g(e_1,\ldots,e_n))
\\
&= \sum_{u_1}\cdots\sum_{u_n} P_u(u_1,\ldots,u_n) \cdot C(z= g(u_1,\ldots,u_n))\cdot
\\
&\quad P_u(u_1,\ldots,u_n)= 
\sum_{x_1}\cdots\sum_{x_n} P_x(x_1,\ldots,x_n) \cdot C(u_1 = e_1)\cdots  C(u_n = e_n)
\enddot
\end{align*}
This rule extends the program with an extra probability function $P_{u}$. 
We assume that the programs do not have unrestricted recursion 
and therefore
we will only generate a bounded number of extra probability functions.

\paragraph{Conditional expressions}
For conditional expressions we use the following rule
\begin{align*}
&\sum_{x_1}\cdots\sum_{x_n}
 P_x(x_1,\ldots,x_n) \cdot C(z= 
 \kw{if}(b(x_1,\ldots,x_n)) \kw{then} g(x_1,\ldots,x_n) \kw{else} h(x_1,\ldots,x_n) )
\\& =
\sum_{x_1}\cdots\sum_{x_n} P_x(x_1,\ldots,x_n) \cdot 
\\&\quad\quad
(C(b(x_1,\ldots,x_n)) \cdot c(z=g(x_1,\ldots,x_n)) +
 C(\neg b(x_1,\ldots,x_n)) \cdot c(z=h(x_1,\ldots,x_n) )\, )
\enddot
\end{align*}

\paragraph{Unfolding primitive recursion}
For primitive recursion we collect the probability of a given result being returned for any number of recursive calls. 
The condition may never evaluate to true for a certain input (non-termination), 
and in that situation the sum of output probabilities will be less than 1.

The recursive functions have the form
\begin{align*}
&f(x_1,\ldots,x_n) = \kw{if}(b(x_1,\ldots,x_n)) \kw{then} g(x_1,\ldots,x_n) \kw{else} f(e_1\ldots,e_n)
\end{align*}
and they should be analyzed for all input probability distributions we detect at calls to these functions.

The transformation for the primitive recursive form is
\begin{align*}
&\sum_{x_1}\cdots\sum_{x_n}
P_x(x_1,\ldots,x_n) \cdot C\left(z= 
\kw{if}(b(x_1,\ldots,x_n)) \kw{then} g(x_1,\ldots,x_n) \kw{else} 
 f(e_1\ldots,e_n)
\right)
\\
&= \sum_{x_1}\cdots\sum_{x_n}
P_x(x_1,\ldots,x_n) 
\sum_{i=0}^\infty
   \prod^{i-1}_{j=0} 
   C(\neg b(h(j,x_1,\ldots,x_n))) \cdot C( b(h(i,x_1,\ldots,x_n))) 
\\&\qquad\qquad   \qquad\qquad\qquad\qquad\qquad
   \cdot C(z = g(h(i,x_1,\ldots,x_n))) 
\end{align*}
\mbox{where}
\begin{align*}
&
h(i,x_1,\ldots,x_n) = \kw{if}(i=0) \kw{then} \bc{x_1,\ldots,x_n} \kw{else} h(i-1,e_1,\ldots,e_n)
\enddot
\mspaceend
\end{align*}
In the transformed expression we introduce two variables: $i$ that represents the number of recursive calls, and $j$ that represents all previous recursions for the $i$ under investigation (when $i$ is 0 the 
term $\prod^{i-1}_{j=0} C(\neg b(h(j,x_1,\ldots,x_n)))$ evaluates to 1). 
The new function $h(i,x_1,\ldots,x_n)$ describes the evaluation of the expressions $\bc{e_1,\ldots,e_n}$, $i$ times.
Only when the $i$th condition is {\it true}  and all previous conditions are {\it false} 
can the expression evaluate to a probability above 0.

\subsection{Symbolic summation}
In the previous phase we unfolded calls to functions in the original program. 
The aim of this phase is to use algebraic transformation techniques to remove summations. 
The methods we use are similar to the transformations used in 
worst case execution time system for solving recurrence equations
\cite{conf/birthday/Rosendahl02,conf/iclp/Lopez-GarciaDB10} 
or symbolic summation techniques in loop bound computations \cite{conf/ershov/KnoopKZ11}. 
Some of the central transformation rules we apply in this phase are listed below.
In the following transformations the expressions $e_1$ and $e_2$ are assumed not to contain the summation variable $x$.
\begin{align*}
\mspacebegin
&  \sum_x C(x = e_1) \cdot f(x)
   =  f(e_1)\\
&  \sum_x C(e_1 \leq x \leq e_2)
   =  (e_2 - e_1 + 1) \cdot C(e_1 \leq e_2)\\
&  \sum_x x \cdot C(e_1 \leq x \leq  e_2)
   =
     \left( \frac{e_2 \cdot (e_2 + 1)}{2}  - \frac{e_1 \cdot (e_1 - 1)}{2} \right)\!
    \cdot\! C(e_1 \leq e_2) 
\enddot
\mspaceend
\end{align*}
One could also use computer algebra systems in the 
reduction process but some of the rules are quite specific to the way we handle
the boundaries of summations with the special constraint function. There are a number of rules to combine products of constraint functions and to split intervals into separate expressions.
\begin{align*}
\mspacebegin
&  C(e_1 \leq x \leq  e_2) \cdot C(e_3 \leq x \leq  e_4) 
   =
     C(\max(e_1,e_3)\leq x \leq \min(e_2,e_4) )  \\[10pt]
& C(\max(e_1, e_2)\leq e_3) 
  =
    C(e_1 > e_2)\! \cdot\! C(e_1 \leq e_3)
     + C(e_1 \leq e_2)\! \cdot \!C(e_2 \leq e_3)  
\enddot
\mspaceend
\end{align*}
There are similar rules for removing the minimum function 
and for isolating variables in constraints.

There are also rules for symbolic summation of certain infinite summations.
If $a$ is an expression where $0 < a < 1$ then we can simplify the expression as follows:
\begin{align*}
\mspacebegin
&\displaystyle
\sum_x C(x \geq 0) \cdot  a^x =
\frac{1}{(1-a)}
\\
&\displaystyle
\sum_x C(x \geq 0) \cdot x\cdot a^x =
\frac{1}{(1-a)^2} - \frac{1}{(1-a)}
\enddot
\mspaceend
\end{align*}
This rule is useful when some of the input to the program follows a geometric distribution.
\begin{align*}
&\displaystyle
P_x(x,n) = C(x \geq 0)\cdot \frac{1}{n}\cdot{\left(1-\frac{1}{n}\right)}^x
\enddot
\mspaceend
\end{align*}
Finally, there are rules for removing finite products and performing standard algebraic simplification transformations.

\paragraph{Max example}  
As a  small example, let us look 
at the simple non-recursive program \verb+max+, which given two values return the largest. 
It is chosen because of its simplicity while still producing a non-uniform output distribution.
The program is defined as
\begin{quote}\vspace{3pt}
\begin{verbatim}	
 max(x,y) = if (x>y) then x else y
\end{verbatim}\vspace{3pt}
\end{quote}
The input values are independent and they follow a uniform distribution from 1 to $n$:
\begin{align*}
\mspacebegin
&
P_x(x) =\frac{1}{n} \cdot C(1 \leq x \leq n) \quad \mathrm{ and } \quad 
P_y(y) =\frac{1}{n} \cdot C(1 \leq y \leq n)
\enddot
\mspaceend
\end{align*}
The output probability program is constructed and simplified 
using transformation rules for conditional expressions and symbolic summation.
\begin{align*}
& P_{\texttt{max}}(z) 
= \sum_x \sum_y  P_x(x)\cdot P_y(y)
   \cdot  C(z=\texttt{if (}x>y)\texttt{ then }x\texttt{ else }y)   
\\& =\displaystyle
 \frac{1}{n^2} \cdot \Big(  \quad 
     \sum_y \big(  C(1 \leq z \leq n) \cdot C(1 \leq y \leq n) 
       \cdot C(y\leq(z-1)) 
     \, \big) \Big. 
\\&\displaystyle \Big. \qquad \qquad 
    +  \sum_x \big( C(1 \leq x \leq n) 
        \cdot  C(1 \leq z \leq n) \cdot C(x\leq z)
    \,\big) \Big)  
\\
&\displaystyle = \frac{1}{n^2} \cdot (2z-1) \cdot C(1 \leq z \leq n) 
\enddot
\end{align*}
The output probability program takes the output value ($ z$) as input and uses the range of input variables ($n$) as an implicit parameter.

\paragraph{Add example}
The recursive addition function was used as an example in the introduction.
We shall see how the original program is inserted into the probability formula, expanded and reduced to 
a closed form function expressing the probability distribution for the output.
Recall the program:
\begin{quote}\vspace{3pt}
\begin{verbatim}
 add(x,y) = if (x=0) then y else add(x-1,y+1)
\end{verbatim}\vspace{3pt}
\end{quote}
and that we assume independence between the input variables and 
that
both input variables \texttt{x} and \texttt{y} have a uniform distribution from 1 to a number $n$. 
The output probability program is constructed using the rule for primitive recursion.
\begin{align*}
\mspacebegin
&P_{\texttt{add}}(z) 
 = \sum_x \sum_y P_x(x) \cdot P_y(y) \cdot 
 \sum_{i=0} \prod_{j=0}^{i-1} 
  C(\neg b(h(j,x,y)))\cdot
  C(b(h(i,x,y)))
  \cdot C(z=g(h(i,x,y))
\end{align*}
\vspace*{-10pt}
\\\noindent
where
\vspace*{-10pt}
\begin{align*}
&b(x,y) = x=0&\\
&g(x,y) = y\\
&h(i,x,y) 
 =\kw{if}(i=0)\kw{then}\bc{x,y}\kw{else}h(i-1,x-1,y+1)
\\&\hphantom{h(i,x,y)}
 =\bc{x-i,y+i}
\end{align*}
The simplification process proceeds as follows.
\begin{align*}
&P_{\texttt{add}}(z) 
   \displaystyle=
\\&\displaystyle\hphantom{=}
    \sum_x \sum_y P_x(x) \cdot P_y(y) \cdot 
    \sum_{i=0} \prod_{j=0}^{i-1} 
    C(\neg (x-j=0))\cdot
    C(x-i=0)\cdot C(z=y+i)
\\[\smallskipamount]&\displaystyle 
  = \sum_x \sum_y P_x(x) \cdot P_y(y) \cdot 
    \sum_{i=0}  C(x=i)\cdot C(z=y+i)
\\[\smallskipamount]&\displaystyle 
  = \sum_x \sum_y P_x(x) \cdot P_y(y) \cdot C(z=y+x)
\\[\smallskipamount]&\displaystyle
 = \sum_y \frac{1}{n} \cdot C(z-n \leq y \leq z-1) \cdot \frac{1}{n} \cdot 
   C(1 \leq y \leq n)  \big) 
\\&\displaystyle
 = \frac{1}{n^2}  \cdot \max(\min(n,z-1) - \max(1,z-n) + 1 , 0)
\\&\displaystyle
 = \frac{1}{n^2}  \cdot \big( C(n<z\leq 2n) \cdot (2n-z+1) 
\quad
 +
\quad
  C(1 \leq z \leq n) \cdot (z-1) \big) 
\enddot
\mspaceend
\end{align*}
The probability program computes a triangular shaped probability distribution with a maximum at $n+1$.

\subsection{Expected value}
When programs produce numeric output values we can use the probability output distribution to compute the expected value or an expected value interval for the output values. The expected value is defined as

\begin{align*}
\mspacebegin
&
E_{\mathrm p} = \sum_{x}  x \cdot P_{\mathrm p}( x )&
\enddot
\mspaceend
\end{align*}
For the \texttt{add} program
this gives
\begin{align*}
\mspacebegin
E_{\texttt{add}}
= \sum_{z=1}^{n} z \cdot \frac{1}{n^2}  \cdot(z-1) 
+  \sum_{z={n+1}}^{2n} z \cdot \frac{1}{n^2}  \cdot (2n-z+1) &&
\mspaceend
\end{align*}
which, of course, can be reduced further.

\section{Composite Types}

In the approach we have presented the base domain is a countable set and not necessarily just numbers. 
We only need to be able to define a probability distribution for values in the domain.

For lists of length $k>0$ where elements are uniformly distributed over the interval $1$ to $n$ we can use the probability function
\begin{align*}
\mspacebegin
&
P_L(L) = 
\frac{1}{n^k}
\cdot C(\textit{length}(L) = k \land 
\forall j:0\leq j\leq k-1 \land 1 \leq hd(tl^j(L)) \leq n)
\enddot
\mspaceend
\end{align*}
We assign the probability ${1}/{n^k}$ to any list of length $k$ where all elements are in the interval from $1$ to $n$.

If we consider the member function for non-empty lists, it can be written as
\begin{quote}
\begin{verbatim}
 member(X,L) = if (tl(L)=[] || hd(L)=X) then hd(L)=X 
               else member(X,tl(L))
\end{verbatim}
\end{quote}
The function will follow the pattern of primitive recursion as described earlier and 
the output probability distribution for the member function is then
\begin{align*}
\mspacebegin
P_{\texttt{member}}(z) =
\sum_X \sum_L P_X(X) \cdot P_L(L) \cdot C(z=\texttt{member}(X,L))
\enddot
\mspaceend
\end{align*}
We can then use the unfolding rules to simplify the expression further.

The lists were here assumed to contain possibly repeating elements in the list. 
We could also use a different probability measure to restrict lists to non-repeating lists of values. 
The example is also analyzed by Wegbreit \cite{journals/cacm/Wegbreit75} where he 
derives the probability as
$1-(1-(1/n)^k)$. It is analyzed under the assumption of non-repeating lists but is
actually the correct result for repeating lists.

His technique is valid for programs where one can safely assume the Markov property (that probabilities of conditions are fixed). 
Wegbreit observes that this is not always true even in very simple cases, e.g. 
in nested conditionals where the outcome of the first condition influences the probability of the outcome for the subsequent condition. 

It should be noted that conditions existing inside a recursive structure often invokes dependencies between variables. 
This occur when there is a \textit{gain of knowledge}: For instance in the \texttt{union} function for two repeating lists; if the head of the list is not in the second list, the likelihood of the next element not being in the second list increases slightly.


\section{Approximation Techniques}
\label{sec:ApproxTec}

The probability distribution program expresses the probability distribution for output values. 
Our aim is to transform it into a closed form but this may not always be possible. 
Failing that, we can instead use approximation techniques to obtain an upper bound for the probability distribution. 
We have referred to this as the over approximation of the probability distribution, $\Ppup$. 

\paragraph{Cumulative distribution functions}
Cumulative probabilities will in some cases be more useful and expressive than probability distributions: Cumulative probabilities can be used in both the discrete and the continuous case, and in some cases approximations can be described more precisely using accumulated probabilities than with ordinary distributions. It tends, however, to be more complex to reduce to closed form and thus may require coarser approximations. The bounding of a cumulative distribution was introduced by Ferson \cite{tech:sand:Ferson2002} as a P-box and can be used to describe imprecise probability distributions.

\begin{definition}[cumulative distribution]
Given a program output probability distribution, $P_{\mathtt{p}}(z)$, the cumulative program output probability distribution, $F_{\mathtt{p}}(z)$, is defined as
\smallskip
\begin{align*} 
&
F_{\mathtt{p}}(z) = \sum_{w \leq z} P_{\mathtt{p}}(w) 
\enddot
\end{align*}
\end{definition}

\begin{definition}[over and under approximation]\label{def:approximationF}
Given a cumulative output probability of a program \mbox{\tt p}, $F_{\mathtt{p}}$, the over approximation, $ \Fup{}_{\mathtt{p}}$, and the under approximation, $ \Fdwn{}_{\mathtt{p}}$, are defined as 
\begin{align*}
\mspacebegin
&   \Fup{}_{\mathtt{p}} :  \forall z . 
    F_{\mathtt{p}}(z) \leq  \Fup{}_{\mathtt{p}}(z) 
\quad\quad
&   \Fdwn{}_{\mathtt{p}} :  \forall z .   \Fdwn{}_{\mathtt{p}}(z) \leq F_{\mathtt{p}}(z)
\mspaceend
 \end{align*}
where for each approximation it must always hold that 
\begin{align*}
\mspacebegin
\forall z . 0 \leq  \Fup{}_{\mathtt{p}}(z) \leq 1 
&\quad\quad
\forall z . 0 \leq  \Fdwn{}_{\mathtt{p}}(z) \leq 1 
\mspaceend
\enddot
 \end{align*}
\end{definition}

When we can deduce that a program may return one of two values, but not which one, then the cumulative probability can be used for a more precise description. 
Such a program could be \texttt{if x = 1 then 1 else (if x = 4 then 4 else (if (unanalyzable) then 2 else 3))} and $1 \leq \mathtt{x} \leq 4$ with the probability distribution $P_x(x) = 1/4\cdot C(1\leq x\leq 4) $. 

\bigskip

\includegraphics[scale=0.8]{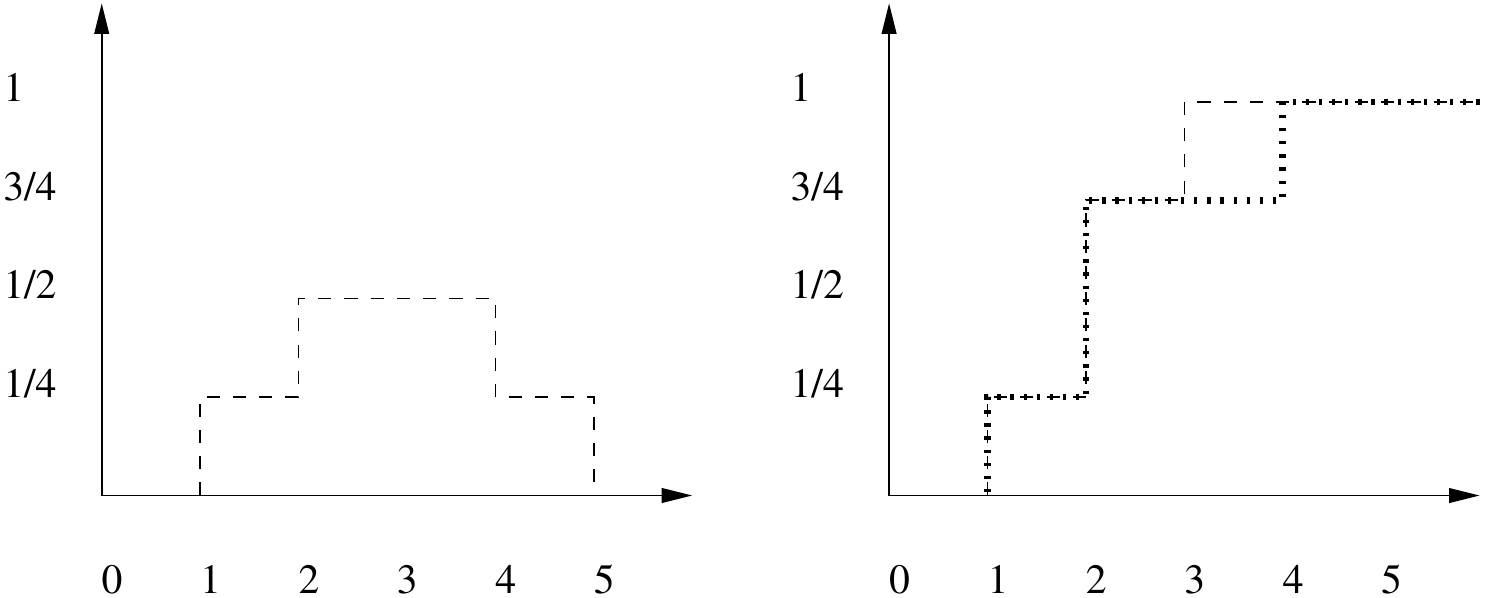}

\medskip

\noindent
Here, the over approximating distribution function will assign  $1/2$ for both 2 and 3. 
In contrast, the over approximating cumulative distribution can express that if the program-output is not 2 it must be 3. 

The distributions
$ \Ppdwn{}$ and $ \Ppup{}$ can be used to derive $ \Fdwn{}_{\mathtt{p}}$ and $ \Fup{}_{\mathtt{p}}$. 
However, these may not be as precise as cumulative distributions derived directly.


When approximating cumulative probability distributions the techniques are different from probability mass functions. 
Instead one may use copulas 
\cite{journals/jec/BernatBN05} to over and under approximate dependencies between subexpressions. 
Copulas are based on the theory of  comonotonicity
\cite{journals:ima:DhaeneDGKV02} for distributions that may depend on a common (possibly unknown) random variable.


\section{Related Work}
\label{sec:relatedWork}

Probabilistic analysis is related to the analysis of probabilistic programs. 
Probabilistic analysis is analysis of programs with a normal semantics where the input variables are interpreted over probability distributions. 
Analysis of probabilistic programs analyzes programs with probabilistic semantics where the values of the input variables are unknown (e.g. 
flow analysis \cite{conf:pfis:Pierro2013}).

In probabilistic analysis it is important to determine how variables depend on each other, but already in 1976 Denning proposed a flow analysis for revealing whether variables depend on each other \cite{journals/cacm/Denning76}. 
This was presented in the field of secure flow analysis.
Denning introduced a lattice-based analysis where she, given the name of a variable, that should be kept secret, deducted which other variables those should be kept secret in order to avoid leaking information. 
In 1996, Denning's method was refined by Volpano {\etal} into a type system and for the first time, it was proven sound \cite{journals/jcs/VolpanoIS96}. 

Reasoning about probabilistic semantics is a closely related area to probabilistic analysis, as they both work with nested probabilistic influence. 
The probabilistic analysis work on standard semantic and analyze it using input probability distributions, where a probabilistic semantics allow for random assignments and probabilistic choices 
\cite{journals/jcss/Kozen81} 
and is normally analyzed using an expanded classical analysis or verification method \cite{conf/esop/CousotM12}. 

Probabilistic model checking is an automated technique for formally verifying quantitative properties for systems with probabilistic behaviors. 
It is mainly focused on Markov decision processes, which can model both stochastic and non-deterministic behavior \cite{conf/sfm/ForejtKNP11,conf:alleton:Kwiatkowska2010}.
It differs from probabilistic analysis as it assumes the Markov property.

In 2000, Monniaux applied abstract interpretation to programs with probabilistic semantics and gained safe bounds for worst case analysis 
\cite{conf/sas/Monniaux00}.
Pierro {\etal} introduce a linear mapping structure, a Moore-Penrose pseudo-inverse, instead of a Galois connection. 
They use the linear structures to compare 'closeness' of approximations as an expression using the average approximation error.  
Pierro {\etal} further explores using probabilistic abstract interpretation to calculate the average case analysis 
\cite{conf/birthday/PierroHW06}.
In 2012, Cousot and Monerau gave a general probabilistic abstraction framework
\cite{conf/esop/CousotM12}
and stated that Pierro {\etal}'s method and many other abstraction methods can be expressed in this new framework.  

When analyzing probabilities the main challenge is to maintain the dependencies throughout the program. 
Schellekens defines this as \textit{Randomness preservation} \cite{books/daglib/0020847} (or random bag preservation) which in his (and Gao's \cite{thesis:cork:gao2013}) case enables tracking of certain data structures and their distributions. 
They use special data structures as they find these suitable to derive the average number of basic operations. 
In another approach \cite{journals/cacm/Wegbreit75,journals:ing:PollmanCG09}, tests in programs has been assumed to be independent of previous history, also known as the Markov property (the probability of true is fixed). 
As Wegbreit remarked, this is true only for some programs (e.g. 
linear search for repeating lists) and others, this is not the case (linear search for non-repeating lists). 
The Markov property is the foundation in Markov decision processes which is used in probabilistic model-checking \cite{conf/sfm/ForejtKNP11}. 
Cousot et al. presents a probabilistic abstraction framework where they divide the program semantics into probabilistic behavior and (non-)deterministic behavior. 
They handle  loops by using a probability function describing the probability of entering the loop in the $i$th iteration. 
Monniaux propose another approach for abstracting probabilistic semantics \cite{conf/sas/Monniaux00}; he first lifts a normal semantics to a probabilistic semantics where random generators are allowed and then uses an abstraction to reach a closed form. 
Monniaux's semantic approach uses a backward probabilistic semantics operating on measurable functions. 
This is closely related to the forward probabilistic semantics proposed earlier by Kozen 
\cite{journals/jcss/Kozen81}. 

An alterntive approach to probabilistic analysis is based on symbolic execution of programs with symbolic values
\cite{conf:issta:Geldenhuys2012}. 
Such techniques can also be used on programs with infinitely many execution paths by limiting the analysis to a finite set of paths at the expense of 
tightness of probability intervals \cite{conf:pldi:SankaranarayananC2013}.


\section{Conclusion}
\label{sec:conclusion}

Probabilistic analysis of programs  has a renewed interest for analyzing programs for energy consumptions. 
Numerous embedded systems and mobile applications are limited by restricted battery life on the hardware.
In this paper we present  a technique for extracting a probability distribution for programs from symbolic distributions of the input.
It is a static transformation based method which can analyze a first order language with primitive recursion.
From the original program and an input probability distribution we generate an output probability distribution program, and transform this program into closed form. 
We present the essential transformation rules for unfolding calls to the original program and removing infinite sums.
The transformed program is then analyzed and approximated using program transformation techniques.
The core elements of the analysis have been implemented in a prototype system with the aim of using it to improve energy efficiency of systems. 
The central challenges of approximating in a probabilistic setting are discussed and we describe some advantages of using cumulative distributions along with copulas to achieve a tighter approximation.

\paragraph{Acknowledgements} This work has benefitted from numerous discussions with Pedro L\'opez-Garc\'ia, Alejandro Serrano Mena and other colleagues in Madrid, Bristol and Roskilde.


\bibliographystyle{eptcs}
\bibliography{dir-ref}
\end{document}